\documentclass[11pt, a4paper]{article}

\PassOptionsToPackage{onlymath}{sansmathfonts}
\AddToHook{package/sansmathfonts/before}{\fontencoding{OT1}\selectfont}
\AddToHook{package/sansmathfonts/after}{\fontencoding{T2A}\selectfont}

\ifcsname dgB@loaded\endcsname\expandafter\endinput\fi
\expandafter\let\csname dgB@loaded\endcsname\empty

\usepackage{ifthen}
\usepackage{etoolbox}%

\makeatletter
\@ifundefined{ifdelim_grow}{%
\newboolean{delim_grow}\setboolean{delim_grow}{true}%
}{}
\@ifundefined{ifare_slides}{%
\newboolean{are_slides}%
\setboolean{are_slides}{false}%
}{}
\@ifundefined{ifdg_impose_style}{%
\newboolean{dg_impose_style}\setboolean{dg_impose_style}{true}%
}{}
\@ifundefined{ifdg_load_tikz}{%
\newboolean{dg_load_tikz}\setboolean{dg_load_tikz}{true}%
}{}
\makeatother

\usepackage{amssymb}
\usepackage{amsmath}%

\usepackage{amsthm}

\usepackage{latexsym}

\ifthenelse{\boolean{are_slides}}{}{\usepackage{enumitem}}

\usepackage{units}%

\usepackage{filemod}%

\usepackage[nolabel]{fnbreak}%

\usepackage{psfrag}%

\makeatletter
\@ifpackageloaded{xcolor}{%
\@ifundefinedcolor{BrickRed}{\input{dvipsnam.def}}{}%
}{\usepackage[dvipsnames]{xcolor}}
\makeatother
\ifthenelse{\boolean{dg_load_tikz}}{\usepackage{tikz}}{}

\makeatletter
\@ifpackageloaded{url}{}{\PassOptionsToPackage{hyphens,spaces,obeyspaces}{url}}
\makeatother
\usepackage{url}%

\usepackage{graphicx}

\usepackage{eso-pic}%

\usepackage{marginnote}%

\usepackage{caption}%

\usepackage{calc}%

\usepackage{soul}%
\setul{1.25pt}{.16pt}%

\usepackage{bm}%

\usepackage{newunicodechar}%
\newunicodechar{⇒}{\ensuremath{\Rightarrow}}
\newunicodechar{ꞏ}{\scalebox{0.75}{\ensuremath{\bullet}}}

\ifthenelse{\boolean{dg_impose_style}}{%
\usepackage{tempora}%

\usepackage[T1,OT2,T2A]{fontenc}

\usepackage[utf8]{inputenc}

\usepackage[shorthands=off,greek,czech,ukrainian,english]{babel}

\DeclareTextSymbolDefault{\CYRYAT}{OT2}
\DeclareTextSymbolDefault{\cyryat}{OT2}
\DeclareTextSymbolDefault{\CYRFITA}{OT2}
\DeclareTextSymbolDefault{\cyrfita}{OT2}
\DeclareTextSymbolDefault{\CYRIZH}{OT2}
\DeclareTextSymbolDefault{\cyrizh}{OT2}

\fontencoding{OT1}\selectfont%
\usepackage[onlymath]{sansmathfonts}
\fontencoding{T2A}\selectfont

}{}%

\usepackage{microtype}%

\usepackage{mathrsfs}%

\usepackage{calligra}%

\usepackage{pdfrender}%

\usepackage{hyperref}%
\setpdflinkmargin{0.75pt}%
\ifthenelse{\boolean{dg_impose_style}}{%
\hypersetup{
colorlinks,
linkcolor={red!55!black},
citecolor={blue!55!black},
urlcolor={blue!35!black}
}
}{}

\ifthenelse{\boolean{are_slides}}{
\usepackage{beamerthemesplit}

\usepackage{ragged2e}%

\usetheme[secheader]{Boadilla}

\hypersetup{hidelinks}%

\usenavigationsymbolstemplate{}%
}{}%

\ifthenelse{\boolean{are_slides}}{
\newcommand{\MyChangeMathMargins}{%
\setlength{\abovedisplayskip}{\abovedisplayshortskip + 3pt}%
\setlength{\belowdisplayskip}{\abovedisplayshortskip + 4pt}%
}

\setul{1.5pt}{0.65pt}%
}{
\newcommand{\MyChangeMathMargins}{%
\setlength{\abovedisplayskip}{\abovedisplayshortskip + 4pt}%
\setlength{\belowdisplayskip}{\abovedisplayshortskip + 5pt}%
}

\ifthenelse{\boolean{dg_impose_style}}{%
\setlength\textheight{9.5in}

\setlength\topmargin{\dimexpr(\paperheight-\textheight
-\headheight-\headsep-\footskip)/2 - 1in\relax}

\setlength\textwidth{6.2in}

\setlength\oddsidemargin{\dimexpr(\paperwidth-\textwidth)/2 - 1in\relax}
\setlength\evensidemargin{\oddsidemargin}

\pdfpageheight=\paperheight
\pdfpagewidth=\paperwidth
}{}%
}%

\makeatletter

\newcommand{\NewcommandThreeArgsTwoOpt}[6][*]{%
\ifthenelse{\equal{}{#1}}%
{\let\comm@TMP\DeclareDocumentCommand}%
{\let\comm@TMP\NewDocumentCommand}%
\expandafter\def\csname\expandafter\@gobble\string#2@nopresq@nopostsq\endcsname##1##2{#3}%
\expandafter\def\csname\expandafter\@gobble\string#2@presq@nopostsq\endcsname##1##2{#4}%
\expandafter\def\csname\expandafter\@gobble\string#2@nopresq@postsq\endcsname##1##2##3{#5}%
\expandafter\def\csname\expandafter\@gobble\string#2@presq@postsq\endcsname##1##2##3{#6}%
\comm@TMP{#2}{o m o}{%
\IfNoValueTF{##1}{%
\IfNoValueTF{##3}%
{\csname\expandafter\@gobble\string#2@nopresq@nopostsq\endcsname{}{##2}}%
{\csname\expandafter\@gobble\string#2@nopresq@postsq\endcsname{}{##2}{##3}}%
}{%
\IfNoValueTF{##3}%
{\csname\expandafter\@gobble\string#2@presq@nopostsq\endcsname{##1}{##2}}%
{\csname\expandafter\@gobble\string#2@presq@postsq\endcsname{##1}{##2}{##3}}%
}%
}%
}

\newcommand{\NewcommandTwoArgsLastOpt}[4][*]{%
\ifthenelse{\equal{}{#1}}%
{\let\comm@TMP\DeclareDocumentCommand}%
{\let\comm@TMP\NewDocumentCommand}%
\expandafter\def\csname\expandafter\@gobble\string#2@nopostsq\endcsname##1{#3}%
\expandafter\def\csname\expandafter\@gobble\string#2@postsq\endcsname##1##2{#4}%
\comm@TMP{#2}{m o}{%
\IfNoValueTF{##2}%
{\csname\expandafter\@gobble\string#2@nopostsq\endcsname{##1}}%
{\csname\expandafter\@gobble\string#2@postsq\endcsname{##1}{##2}}%
}%
}

\newcommand{\NewcommandFourArgsTwoOpt}[6][*]{%
\ifthenelse{\equal{}{#1}}%
{\let\comm@TMP\DeclareDocumentCommand}%
{\let\comm@TMP\NewDocumentCommand}%
\expandafter\def\csname\expandafter\@gobble\string#2@nopresq@nopostsq\endcsname##1##2##3{#3}%
\expandafter\def\csname\expandafter\@gobble\string#2@presq@nopostsq\endcsname##1##2##3{#4}%
\expandafter\def\csname\expandafter\@gobble\string#2@nopresq@postsq\endcsname##1##2##3##4{#5}%
\expandafter\def\csname\expandafter\@gobble\string#2@presq@postsq\endcsname##1##2##3##4{#6}%
\comm@TMP{#2}{o m m o}{%
\IfNoValueTF{##1}{%
\IfNoValueTF{##4}%
{\csname\expandafter\@gobble\string#2@nopresq@nopostsq\endcsname{}{##2}{##3}}%
{\csname\expandafter\@gobble\string#2@nopresq@postsq\endcsname{}{##2}{##3}{##4}}%
}{%
\IfNoValueTF{##4}%
{\csname\expandafter\@gobble\string#2@presq@nopostsq\endcsname{##1}{##2}{##3}}%
{\csname\expandafter\@gobble\string#2@presq@postsq\endcsname{##1}{##2}{##3}{##4}}%
}%
}%
}

\newcommand{\NewcommandFourArgsTwoOptDefa}[5][*]{%
\ifthenelse{\equal{}{#1}}%
{\let\comm@TMP\DeclareDocumentCommand}%
{\let\comm@TMP\NewDocumentCommand}%
\comm@TMP{#2}{O{#3} m m O{#4}}{#5}%
}

\newcommand{\ifndef}[2]{\@ifundefined{#1}{#2}{}}

\newcommand{\IfMathMode}[2]{\ifmmode{#1}\else{#2}\fi}

\DeclareRobustCommand\bk{\@ifnextchar[%
{\lf\langle \bk@bra{}}%
{\@ifnextchar<%
{\lf.\bk@op}
{\lf| \bk@ket{}}%
}%
}

\def\bk@bra#1[#2]{\def\bk@tmp{#2}\ifx\bk@tmp\empty{#1}\else{#2}\fi\@ifnextchar[%
{\md|\hspace{-1.5pt}\md\langle \bk@bra{\def\bk@tmp{#2}\ifx\bk@tmp\empty{#1}\else{#2}\fi}}%
{\@ifnextchar\bgroup%
{\md| \bk@ket{\def\bk@tmp{#2}\ifx\bk@tmp\empty{#1}\else{#2}\fi}}%
{\@ifnextchar<%
{\md| \bk@op}%
{\rt|}%
}%
}%
}

\def\bk@ket#1#2{\def\bk@tmp{#2}\ifx\bk@tmp\empty{#1}\else{#2}\fi\@ifnextchar[%
{\md\rangle\hspace{-1.5pt}\md\langle \bk@bra{\def\bk@tmp{#2}\ifx\bk@tmp\empty{#1}\else{#2}\fi}}%
{\@ifnextchar\bgroup%
{\md\rangle\hspace{-1.5pt}\md| \bk@ket{\def\bk@tmp{#2}\ifx\bk@tmp\empty{#1}\else{#2}\fi}}%
{\@ifnextchar<%
{\md\rangle \bk@op}%
{\rt\rangle}%
}%
}%
}

\def\bk@op<#1>{#1 \@ifnextchar[%
{\md\langle \bk@bra{}}%
{\@ifnextchar\bgroup%
{\md| \bk@ket{}}%
{\@ifnextchar<%
{\tm \bk@op}%
{\rt.}%
}%
}%
}

\newcommand\Cases[1][0pt]{%
\def\Case@skipam{#1}%
\left\{\!\!\!\begin{array}{ll}\Cases@continue%
}

\def\Cases@continue#1#2{\@ifnextchar\bgroup%
{#1 &\txt{#2}\\[\Case@skipam] \Cases@continue}%
{#1 &\txt{#2}\end{array}\Cases@end}%
}

\def\Cases@end{\@ifnextchar[%
{\Cases@end@postsq}%
{\right.}
}

\def\Cases@end@postsq[#1]{\ifthenelse{\equal{\}}{#1}}
{\!\!\right\}}%
{\right.}
}

\newcommand{\MyMakeTheoMacros}[3]{
\expandafter\newcommand\csname\expandafter\@gobble\string#2NostarNoname@DGaux\endcsname[2][]
{\ifthenelse{\equal{}{##1}}%
{\begin{#1}~##2 \end{#1}}%
{\begin{#1}\label{##1}~##2\end{#1}}%
}
\expandafter\newcommand\csname\expandafter\@gobble\string#2StarNoname@DGaux\endcsname[1]
{\begin{#1*}~##1 \end{#1*}}
\newcommand#2{\expandafter\@ifstar%
\expandafter{\csname\expandafter\@gobble\string#2StarNoname@DGaux\endcsname}%
{\csname\expandafter\@gobble\string#2NostarNoname@DGaux\endcsname}%
}

\expandafter\newcommand\csname\expandafter\@gobble\string#2NostarName@DGaux\endcsname[3][]
{\ifthenelse{\equal{}{##1}}%
{\begin{#1}[\e{##2}]~##3 \end{#1}}%
{\begin{#1}[\e{##2}]\label{##1}~##3\end{#1}}%
}
\expandafter\newcommand\csname\expandafter\@gobble\string#2StarName@DGaux\endcsname[2]
{\begin{#1*}[\e{##1}]~##2 \end{#1*}}
\newcommand#3{\expandafter\@ifstar%
\expandafter{\csname\expandafter\@gobble\string#2StarName@DGaux\endcsname}
{\csname\expandafter\@gobble\string#2NostarName@DGaux\endcsname}%
}
}

\DeclareRobustCommand\dgref[2]{\@ifnextchar[%
{#2\dgref@presq}%
{#1\dgref@nopresq}%
}

\def\dgref@presq[#1]{\@ifnextchar[%
{\ref{#1}, \dgref@presq}
{\ref{#1} and~\dgref@nopresq}
}

\def\dgref@nopresq#1{\ref{#1}}

\DeclareRobustCommand\bref{\@ifnextchar[%
{\bref@presq}%
{\bref@nopresq}%
}

\def\bref@presq[#1]{\@ifnextchar[%
{(\ref{#1}), \bref@presq}
{(\ref{#1}) and~\bref@nopresq}
}

\def\bref@nopresq#1{(\ref{#1})}

\makeatother

\newcommand{\NewELike}[2]{
\NewcommandThreeArgsTwoOpt[]{#1}
{#2\lf[##2\rt]}
{#2_{##1}\lf[##2\rt]}
{#2\lf[##2\md|{\mathstrut}##3\rt]}
{#2_{##1}\lf[##2\md|{\mathstrut}##3\rt]}
}

\NewELike{\PR}{\mathop{\pmb{Pr}}}

\NewELike{\E}{\mathop{\pmb{E}}}

\newcommand{\NewMinLike}[2]{
\NewcommandThreeArgsTwoOpt[]{#1}
{\mathop{#2}\lf\{##2\rt\}}
{\mathop{#2}_{##1}\lf\{##2\rt\}}
{\mathop{#2}\lf\{##2\md|{\ud}##3\rt\}}
{\mathop{#2}_{##1}\lf\{##2\md|{\ud}##3\rt\}}
}

\NewMinLike{\Max}{max}

\newcommand{\NewOp}[2]{
\newcommand{#1}[2][]{\ensuremath{\operatorname{#2}##1\ifthenelse{\equal{}{##2}}{}{\lf(##2\rt)}}}
}

\newcommand{\NewOpI}[2]{\NewOp{#1}{\mathit{#2}}}

\NewOp{\Dim}{dim}
\NewOp{\rank}{rank}

\NewOpI{\asO}{O}
\NewOpI{\asOm}{\Omega}
\NewOpI{\asT}{\Theta}

\newcommand{\NewId}[2]{
\newcommand{#1}[1][]{\ensuremath{\ifthenelse{\equal{}{##1}}{\operatorname{\mathit{#2}}}{\operatorname{\mathit{#2##1}}}}}
}

\newcommand{\NewIdR}[2]{
\newcommand{#1}[1][]{\ensuremath{\ifthenelse{\equal{}{##1}}{\operatorname{\mathrm{#2}}}{\operatorname{\mathrm{#2##1}}}}}
}

\newcommand{\NewIdSm}[2]{\NewId{#1}%
{\mathchoice{#2}{#2}{\scriptscriptstyle{#2}}{\scalebox{0.75}{$\scriptscriptstyle{#2}$}}}%
}

\NewIdSm{\tp}{\top}
\NewIdSm{\bt}{\bot}

\newcommand{\NewM}[2]{
\newcommand{#1}{\ensuremath{#2}}
}

\NewM{\OI}{\set{0,1}}
\NewM{\II}{\set{-1,1}}
\NewM{\TF}{\set{\tp,\bt}}

\newcommand{\MyMakeRefMacros}[3]{\newcommand{#1}{\dgref{#2}{#3}}}

\ifndef{theorem}{}
\ifndef{theorem*}{\newtheorem*{theorem*}{Theorem}}

\MyMakeTheoMacros{theorem}{\theo}{\ntheo}

\MyMakeRefMacros{\theoref}{Theorem~}{Theorems~}

\ifndef{definition}{}
\ifndef{definition*}{\newtheorem*{definition*}{Definition}}

\MyMakeTheoMacros{definition}{\defi}{\ndefi}

\MyMakeRefMacros{\defiref}{Definition~}{Definitions~}

\ifndef{claim}{}
\ifndef{claim*}{\newtheorem*{claim*}{Claim}}

\MyMakeTheoMacros{claim}{\clm}{\nclm}

\MyMakeRefMacros{\clmref}{Claim~}{Claims~}

\ifndef{lemma}{}
\ifndef{lemma*}{\newtheorem*{lemma*}{Lemma}}

\MyMakeTheoMacros{lemma}{\lem}{\nlem}

\MyMakeRefMacros{\lemref}{Lemma~}{Lemmas~}

\ifndef{corollary}{}
\ifndef{corollary*}{\newtheorem*{corollary*}{Corollary}}

\MyMakeTheoMacros{corollary}{\crl}{\ncrl}

\MyMakeRefMacros{\crlref}{Corollary~}{Corollaries~}

\ifndef{fact}{}
\ifndef{fact*}{\newtheorem*{fact*}{Fact}}

\MyMakeTheoMacros{fact}{\fct}{\nfct}

\MyMakeRefMacros{\fctref}{Fact~}{Facts~}

\ifndef{question}{}
\ifndef{question*}{\newtheorem*{question*}{Question}}

\MyMakeTheoMacros{question}{\quest}{\nquest}

\MyMakeRefMacros{\questref}{Question~}{Questions~}

\ifndef{construction}{}
\ifndef{construction*}{\newtheorem*{construction*}{Construction}}

\MyMakeTheoMacros{construction}{\constr}{\nconstr}

\MyMakeRefMacros{\constrref}{Construction~}{Constructions~}

\renewcommand{\qedsymbol}{$\blacksquare$}

\DeclareRobustCommand{\prflabel}[1]{%
\ifstrempty{#1}%
{\renewcommand{\qedsymbol}{$\blacksquare$}}%
{\renewcommand{\qedsymbol}{$\blacksquare_{\mbox{\it{\scriptsize{#1}}}}$}}%
}
\newcommand{\prfstart}[1][]{\ifthenelse{\equal{}{#1}}%
{\begin{proof}[\underline{\proofname}]\prflabel{}}%
{\begin{proof}[\underline{\proofname\ #1}]\prflabel{#1}}%
}

\newcommand{\prfend}[1][*]{%
\ifthenelse{\equal{*}{#1}}{}{\prflabel{#1}}%
\end{proof}%
}

\newcommand{\abstart}{\begin{abstract}}
\newcommand{\abend}{\end{abstract}}

\newcommand{\NewSectLike}[3]{
\NewcommandThreeArgsTwoOpt{#1}
{\ifthenelse{\equal{*}{##2}}{#2*}{#2{##2}}}
{#2{##2}\label{##1}}
{#2[##3]{##2}#3{##3}}
{#2[##3]{##2}\label{##1}#3{##3}}
}

\NewSectLike{\sect}{\section}{\sectionmark}

\NewSectLike{\ssect}{\subsection}{\subsectionmark}

\NewSectLike{\sssect}{\subsubsection}{\subsubsectionmark}

\MyMakeRefMacros{\sref}{Section~}{Sections~}
\MyMakeRefMacros{\ssref}{Subsection~}{Subsections~}

\newcommand{\toc}[3][]{%
\ifthenelse{\equal{chapter}{#2}}{\newpage}{}%
\ifthenelse{\equal{part}{#2}}{\newpage}{}%
\ifthenelse{\equal{}{#1}}%
{\phantomsection\addcontentsline{toc}{#2}{#3}}%
{\phantomsection\refstepcounter{#2}\label{#1}\addcontentsline{toc}{#2}{#3}}%
}

\newcommand{\toct}[1]{\toc{section}{#1}}

\makeatletter
\newif\ifdg@bibtoc
\newcommand{\bibtoc}{%
\ifdg@bibtoc\else
\apptocmd{\thebibliography}{\toct{\refname}}%
{\dg@bibtoctrue}%
{\PackageError{dgB}{Cannot add the bibliography TOC entry}%
{Call \string\bibtoc\space after bibliography-related packages.}}%
\fi
}
\makeatother

\newcommand{\bibstart}[1][99]{\bibtoc\begin{thebibliography}{#1}}
\newcommand{\bibend}{\end{thebibliography}}

\MyMakeRefMacros{\fnref}{Footnote~}{Footnotes~}

\makeatletter
\newif\ifdgmath@centered
\newif\ifdgmath@multipage
\newif\ifdgmath@inline
\newif\ifdgmath@valid

\newcommand{\dgmath@error}[1]{%
\PackageError{dgB}{#1}{Use C and P as display options, I alone for inline math, or one equation label.}%
\dgmath@validfalse
}
\newcommand{\dgmath@option}[1]{%
\ifstrempty{#1}{}{%
\ifstrequal{#1}{C}{\dgmath@centeredtrue}{%
\ifstrequal{#1}{P}{\dgmath@multipagetrue}{%
\ifstrequal{#1}{I}{\dgmath@inlinetrue}{%
\ifdefempty{\dgmath@label}{\def\dgmath@label{#1}}%
{\dgmath@error{At most one equation label is allowed}}%
}%
}%
}%
}%
}
\newcommand{\dgmath@renderalign}[2]{\begin{align*} #2 \end{align*}}
\newcommand{\dgmath@rendercenter}[2]{\begin{gather*} #2 \end{gather*}}
\newcommand{\dgmath@renderpagealign}[2]{%
\allowdisplaybreaks\begin{align*} #2 \end{align*}\interdisplaylinepenalty=10000}
\newcommand{\dgmath@renderpagecenter}[2]{%
\allowdisplaybreaks\begin{gather*} #2 \end{gather*}\interdisplaylinepenalty=10000}
\newcommand{\dgmath@renderinline}[2]{$#2$}
\newcommand{\dgmath@renderlabelalign}[2]{%
\begin{align}\begin{split}\label{#1} #2 \end{split}\end{align}}
\newcommand{\dgmath@renderlabelcenter}[2]{%
\begin{equation}\label{#1}\begin{gathered} #2 \end{gathered}\end{equation}}
\newcommand{\dgmath@renderskip}[2]{}

\NewDocumentCommand{\m}{O{} O{} +m}{%
\MyChangeMathMargins
\ifthenelse{\boolean{delim_grow}}{\delimiterfactor=1001}{}%
\begingroup
\dgmath@centeredfalse\dgmath@multipagefalse\dgmath@inlinefalse\dgmath@validtrue
\def\dgmath@label{}%
\dgmath@option{#1}\dgmath@option{#2}%
\ifdgmath@inline
\ifdgmath@centered\dgmath@error{I cannot be combined with C}\fi
\ifdgmath@multipage\dgmath@error{I cannot be combined with P}\fi
\ifdefempty{\dgmath@label}{}{\dgmath@error{I cannot be combined with a label}}%
\fi
\ifdgmath@multipage
\ifdefempty{\dgmath@label}{}{\dgmath@error{Use malabel inside a P display instead of a label option}}%
\fi
\def\dgmath@handler{dgmath@renderskip}%
\ifdgmath@valid
\ifdgmath@inline
\def\dgmath@handler{dgmath@renderinline}%
\else
\ifdefempty{\dgmath@label}{%
\ifdgmath@centered
\ifdgmath@multipage\def\dgmath@handler{dgmath@renderpagecenter}%
\else\def\dgmath@handler{dgmath@rendercenter}\fi
\else
\ifdgmath@multipage\def\dgmath@handler{dgmath@renderpagealign}%
\else\def\dgmath@handler{dgmath@renderalign}\fi
\fi
}{%
\ifdgmath@centered\def\dgmath@handler{dgmath@renderlabelcenter}%
\else\def\dgmath@handler{dgmath@renderlabelalign}\fi
}%
\fi
\fi
\edef\dgmath@next{\endgroup
\expandafter\noexpand\csname\dgmath@handler\endcsname
{\unexpanded\expandafter{\dgmath@label}}{\unexpanded{#3}}}%
\dgmath@next
}

\let\dgampersand\&

\DeclareRobustCommand\&{%
\new@ifnextchar[%
{\dgsep@reposit}%
{\dgampersand}%
}

\def\dgsep@reposit[#1]{\hspace{#1}&\hspace{-#1}}

\makeatother

\mathchardef\breakingcomma\mathcode`\,
{\catcode`,=\active
\gdef,{\breakingcomma\discretionary{}{}{}}
}

\newcommand{\lf}{\mathopen{}\mathclose\bgroup\left}
\newcommand{\rt}{\aftergroup\egroup\right}

\providecommand{\middle}{\big}
\newcommand{\md}{\middle}

\newcommand{\ud}{\vphantom{|_1^1}}

\makeatletter

\newcommand{\NewPar}[4][*]{
\ifthenelse{\equal{}{#1}}
{\def\dg@parfactory{\NewcommandThreeArgsTwoOpt[]}}
{\def\dg@parfactory{\NewcommandThreeArgsTwoOpt}}
\dg@parfactory{#2}
{\lf#3{##2}\rt#4}
{#3{##2}#4}
{\lf#3{##2}\rt#4_{\!##3}}
{#3{##2}#4_{\!##3}}
}
\makeatother

\NewPar{\lrp}{(}{)}
\NewPar{\lra}{\langle}{\rangle}
\NewPar[]{\sz}{|}{|}
\NewPar[]{\norm}{\|}{\|}
\NewPar[]{\ceil}{\lceil}{\rceil}
\NewPar[]{\floor}{\lfloor}{\rfloor}
\NewPar{\round}{\lfloor}{\rceil}

\NewcommandThreeArgsTwoOpt[]{\set}
{\lf\{#2\rt\}}
{}
{\lf\{#2\md|\ud#3\rt\}}
{\lf\{#2\ud#1\ud#3\rt\}}

\definecolor{Red}{rgb}{1,0,0}
\definecolor{DarkRed}{rgb}{0.54,0,0}
\definecolor{DarkGreen}{rgb}{0, 0.52, 0.05}
\definecolor{LightBlue}{rgb}{0.122,0.016,0.855}

\definecolor{dgProbar}{rgb}{0.3,0.3,0.3}%
\definecolor{dgPreBack}{RGB}{209,211,214}%

\DeclareTextFontCommand{\bemph}{\bfseries}
\DeclareTextFontCommand{\ibemph}{\bfseries\em}
\DeclareTextFontCommand{\ttl}{\ttfamily}

\ifthenelse{\boolean{are_slides}}{

\newcommand{\e}{\ei}
\newcommand<>{\ei}[1]{\IfMathMode%
{\only#2{\textcolor{BrickRed}}{#1}}%
{\emph{\only#2{\textcolor{BrickRed}}{#1}}}%
}
\newcommand<>{\eb}[1]{\IfMathMode%
{\only#2{\textcolor{BrickRed}}{\pmb{#1}}}%
{\bemph{\only#2{\textcolor{BrickRed}}{#1}}}%
}
\newcommand<>{\ebi}[1]{\IfMathMode%
{\only#2{\textcolor{BrickRed}}{\pmb{#1}}}%
{\ibemph{\only#2{\textcolor{BrickRed}}{#1}}}%
}
\newcommand<>{\eu}[1]{\IfMathMode%
{\only#2{\textcolor{BrickRed}}{\txt{\ul{#1}}}}%
{\only#2{\textcolor{BrickRed}}{\ul{#1}}}%
}
\newcommand<>{\eui}[1]{\IfMathMode%
{\only#2{\textcolor{BrickRed}}{\txt{\ul{#1}}}}%
{\emph{\only#2{\textcolor{BrickRed}}{\ul{#1}}}}%
}
\newcommand<>{\ebu}[1]{\IfMathMode%
{\only#2{\textcolor{BrickRed}}{\underline{\pmb{#1}}}}%
{\bemph{\only#2{\textcolor{BrickRed}}{\ul{#1}}}}%
}
\newcommand<>{\ebui}[1]{\IfMathMode%
{\only#2{\textcolor{BrickRed}}{\underline{#1}}}%
{\ibemph{\only#2{\textcolor{BrickRed}}{\ul{#1}}}}%
}
}{%
\newcommand{\e}{\emph}
\newcommand{\eb}{\bemph}
\newcommand{\ebi}{\ibemph}
\newcommand{\eu}[1]{\ttl{\ul{#1}}}
\newcommand{\eui}[1]{\emph{\ul{#1}}}
\newcommand{\ebu}[1]{\bemph{\ul{#1}}}
\newcommand{\ebui}[1]{\ibemph{\ul{#1}}}
}%

\newcommand{\txt}[1]{\textrm{#1}}%

\newcommand{\fr}{\frac}

\newcommand{\tm}{\cdot}
\let\sq\relax%
\newcommand{\sq}{\sqrt}

\newcommand{\ds}[1][]
{\ifthenelse{\equal{}{#1}}{\allowbreak\ldots}{#1\allowbreak\ldots#1}}
\newcommand{\dc}{\ds[,]}

\newcommand{\deq}{\stackrel{\textrm{def}}{=}}

\NewcommandFourArgsTwoOptDefa{\overlay}
{0mu}{1}
{
\begingroup
\mathchoice{%
\bgroup
\scalebox{#4}{\ooalign{$\displaystyle#2\mkern#1$\cr
\hidewidth{$\displaystyle\mkern#1#3$}\hidewidth}}
\egroup
}{%
\bgroup
\scalebox{#4}{\ooalign{$\textstyle#2\mkern#1$\cr
\hidewidth{$\textstyle\mkern#1#3$}\hidewidth}}
\egroup
}{%
\bgroup
\scalebox{#4}{\ooalign{$\scriptstyle#2\mkern#1$\cr
\hidewidth{$\scriptstyle\mkern#1#3$}\hidewidth}}
\egroup
}{%
\bgroup
\scalebox{#4}{\ooalign{$\scriptscriptstyle#2\mkern#1$\cr
\hidewidth{$\scriptscriptstyle\mkern#1#3$}\hidewidth}}
\egroup
}
\endgroup
}

\makeatletter
\DeclareRobustCommand{\qto}{%
\begingroup
\m@th
\def\dg@quantumwave{\mathchoice
{\raisebox{.9ex}{$\displaystyle\sim$}}
{\raisebox{.9ex}{$\textstyle\sim$}}
{\raisebox{.9\fontdimen5\scriptfont2}{$\scriptstyle\sim$}}
{\raisebox{.9\fontdimen5\scriptscriptfont2}{$\scriptscriptstyle\sim$}}}%
\mathrel{\vphantom{\dg@quantumwave}\overlay{\rightarrow}
{\mkern-2.25mu\dg@quantumwave\mkern2.25mu}}%
\endgroup
}
\makeatother
\newcommand{\unin}{\mathrel{\overlay[1.25mu]{\subset}{\sim}}}

\newcommand{\sbseq}{\subseteq}

\NewcommandTwoArgsLastOpt{\GF}
{{\mathcal{G\!F}_{#1}}}
{{\mathcal{G\!F}_{#1}^{#2}}}

\def\?{\mskip 0.5\thinmuskip}%

\newcommand{\qd}[1][1]{\hspace*{#1em}\ignorespaces}%

\makeatletter
\ifthenelse{\boolean{are_slides}}{

\NewcommandTwoArgsLastOpt{\ssframe}
{\only<1>{\ssect{#1}}\frametitle{#1}}
{\only<1>{\ssect{#1}[#2]}\frametitle{#1}}

\newdimen\progressbar@width
\newcount\progressbar@mypage
\newcount\progressbar@topage

\setbeamertemplate{footline}{
\color{dgProbar}

}

\setbeamercolor{background canvas}{bg=dgPreBack}
}{}%
\makeatother

\DeclareFontSubstitution{U}{ssmsa}{m}{n}
\usepackage{mathtools}
\usetikzlibrary{arrows.meta,calc}

\numberwithin{theorem}{section}

\renewenvironment{abstract}{%
\small
\begin{center}
{\bfseries\abstractname\vspace{-.5em}\vspace{0pt}}
\end{center}
\list{}{\leftmargin1em\rightmargin\leftmargin}
\item\relax
}{\endlist}

\makeatletter
\let\dg@articlemaketitle\maketitle
\renewcommand{\maketitle}{{\let\rlap\hbox\dg@articlemaketitle}}
\makeatother

\MyMakeTheoMacros{proposition}{\prp}{\nprp}
\MyMakeRefMacros{\prpref}{Proposition~}{Propositions~}
\MyMakeRefMacros{\figref}{Figure~}{Figures~}
\newcommand{\F}{\mathbb F_2}
\NewOp{\diag}{diag}
\newcommand{\Rone}{R^{\rightarrow}}
\newcommand{\Qone}{Q^{\rightarrow}}
\newcommand{\Qsmp}{Q^{\parallel}}
\newcommand{\Rsmp}{R^{\parallel}}
\newcommand{\Rent}{R^{\parallel,\mathrm{ent}}}
\newcommand{\Rentlim}[1]{R^{\parallel,\mathrm{ent}(#1)}}
\newcommand{\one}{\mathbf 1}

\hypersetup{
pdftitle={Asymptotic advantage of quantum one-way and SMP communication for total functions},
pdfauthor={Dmytro Gavinsky}}
\title{Asymptotic advantage of quantum one-way and SMP communication for total functions}
\author{Dmytro Gavinsky\thanks{In 2022 the author changed his first name, as explained on his Internet page.}%
\thanks{This work was developed in collaboration with {\fontencoding{T1}\fontfamily{lmr}\selectfont\textsc{Anna}} -- Anonymous Neural-Network Assistant -- whom the author sincerely thanks.}\\[0.4ex]
{\small Institute of Mathematics of the Czech Academy of Sciences,
\v{Z}itn\'a 25, Praha~1, Czechia}\\[-0.1ex]
{\small IonQ Quantum, Inc., 4505 Campus Drive,
College Park, MD 20740, USA}}
\date{}

\begin{document}
\maketitle \thispagestyle{empty}

\abstart
\noindent We exhibit a total Boolean function on $n$ input bits whose bounded-error quantum simultaneous-message (and hence one-way) communication complexity is in $\asO{\sq n}$, whereas its randomised one-way (and hence simultaneous-message) communication complexity is in $\asOm{n}$.
The same function admits a classical simultaneous-message protocol of cost $\asO{\sq n}$ when the players share $\asO{\sq n}$ EPR pairs.
\abend

\sect[sec:model]{The problem}

All input algebra is over $\F$.
Alice receives $M\in\F^{d\times d}$, Bob receives $u,z\in\F^d$ and we define
\m[C]{
A_M\deq M+M^{\mathsf T},\qd f_M(x)\deq x^{\mathsf T}Mx,\\
F_d(M;u,z)\deq\one\lf[A_Mu=z\ \txt{and}\ f_M(u)=0\rt].
}
The encoded input length is
\m[I]{
n\deq d^2+2d.
}
Note that the matrix $A_M$ is alternating ($u^{\mathsf T}A_Mu \equiv 0$).

In a one-way protocol, Alice sends one message to Bob, who outputs the answer.
In a simultaneous-message (SMP) protocol, Alice and Bob each send one message to a referee, who outputs the answer.
Write $\Rone,\Qone$ for the classical and quantum one-way complexities, and $\Rsmp,\Qsmp$ for their SMP counterparts.
By default, the models have no prior entanglement; the superscript ${}^{\mathrm{ent}}$ allows an arbitrary prior entangled state between Alice and Bob and ${}^{\mathrm{ent}(e)}$ restricts that state to at most $e$ qubits per party (note that shared entanglement does not, in general, turn $\Rsmp$ into $\Qsmp$).
All models allow public randomness.
Local computation and classical encoders and decoders are unrestricted.
All protocols must have error probability at most $1/3$ on any input -- unless the target distribution is specified.
The cost of a protocol is the maximum total transmitted number of (qu)bits.

Trivially, for any function $f$ and integer $k\ge0$,
\m[C]{
\Rone(f)\le\Rsmp(f),\qd \Qone(f)\le\Qsmp(f),\\
\Qone(f)\le\Rone(f),\qd \Qsmp(f)\le\Rsmp(f),\\
\Qone(f)\le k+\Rentlim{k}(f).
}

We establish the following.
\theo[thm:main]
{
\m[C]{
\Qsmp(F_d),\,\Qone(F_d)~\in~\asT{d}=\asT{\sq n},\\
\Rone(F_d),\,\Rsmp(F_d)~\in~\asT{d^2}=\asT{n},\\
\Rentlim{2d}(F_d)~\in~\asO{d},\qd
\Rent(F_d)~\in~\asOm{d}.
}
}

\sect[sec:qua-com]{The quantum complexity}

\ssect{The one-way protocol}

\begin{figure}[ht]
\centering
\resizebox{\linewidth}{!}{%
\begingroup
\definecolor{acceptgreen}{RGB}{20,112,77}
\definecolor{imageblue}{RGB}{45,83,150}
\begin{tikzpicture}[x=1cm,y=1cm,
state/.style={-{Latex[length=2.2mm]},line width=.8pt},
accepted/.style={-{Latex[length=2.4mm]},line width=1.2pt,acceptgreen},
every node/.style={font=\small}]
\path[use as bounding box] (.2,.5) rectangle (15.2,6.6);
\node[acceptgreen] at (7.7,6.25)
{$\txt{accepted component}=\fr{I+P}{2}\psi=\fr{\psi+P\psi}{2}$};
\node at (7.7,5.65) {$\psi=\psi_M(0),\qd P=P_{u,z},\qd u\tm z=0$};

\node at (2.25,4.85) {$z=A_Mu,\qd f_M(u)=0$};
\node at (7.5,4.85) {$z=A_Mu,\qd f_M(u)=1$};
\node at (12.65,4.85) {$z\ne A_Mu$};
\node[font=\footnotesize] at (2.25,4.35) {coincident vectors};
\node[font=\footnotesize] at (7.5,4.35) {opposite vectors};
\node[font=\footnotesize] at (12.65,4.35) {orthogonal vectors};

\coordinate (O1) at (1.3,2.7);
\coordinate (E1) at (3.1,2.7);
\draw[accepted] (O1)--(E1);
\fill (O1) circle (1.1pt);
\fill[acceptgreen] (E1) circle (1.5pt);
\node[above] at (2.2,2.85) {$\psi=P\psi$};
\node[below left,font=\footnotesize] at (O1) {$0$};

\coordinate (O2) at (7.5,2.7);
\coordinate (E2) at (9.3,2.7);
\coordinate (F2) at (5.7,2.7);
\draw[gray!55,densely dashed] (F2)--(E2);
\draw[state] (O2)--(E2);
\draw[state,imageblue] (O2)--(F2);
\fill[acceptgreen] (O2) circle (2pt);
\node[above] at (E2) {$\psi$};
\node[above,imageblue] at (F2) {$P\psi$};
\node[below,acceptgreen] at (O2) {$0$};

\coordinate (O3) at (11.4,2.0);
\coordinate (E3) at (13.2,2.0);
\coordinate (F3) at (11.4,3.8);
\coordinate (M3) at (12.3,2.9);
\draw[acceptgreen!45,densely dotted,line width=.6pt] (O3)--(13.2,3.8);
\draw[gray!60,densely dashed] (E3)--(F3);
\draw[state] (O3)--(E3);
\draw[state,imageblue] (O3)--(F3);
\draw[accepted] (O3)--(M3);
\fill (O3) circle (1.1pt);
\fill[acceptgreen] (M3) circle (1.7pt);
\node[below] at (E3) {$\psi$};
\node[left,imageblue] at (F3) {$P\psi$};
\node[anchor=west,acceptgreen] at (12.65,3.15) {$\fr{\psi+P\psi}{2}$};
\node[below left,font=\footnotesize] at (O3) {$0$};
\draw[gray!65,line width=.4pt]
($(M3)+(-.11,-.11)$)--($(M3)+(0,-.22)$)--($(M3)+(.11,-.11)$);

\node[acceptgreen] at (2.25,.95) {$\PR{+}=1$};
\node[acceptgreen] at (7.5,.95) {$\PR{+}=0$};
\node[acceptgreen] at (12.65,.95) {$\PR{+}=\fr12$};

\end{tikzpicture}
\endgroup%
}
\caption{The accepting component is shown in green.}
\label{fig:oneway-geometry}
\end{figure}
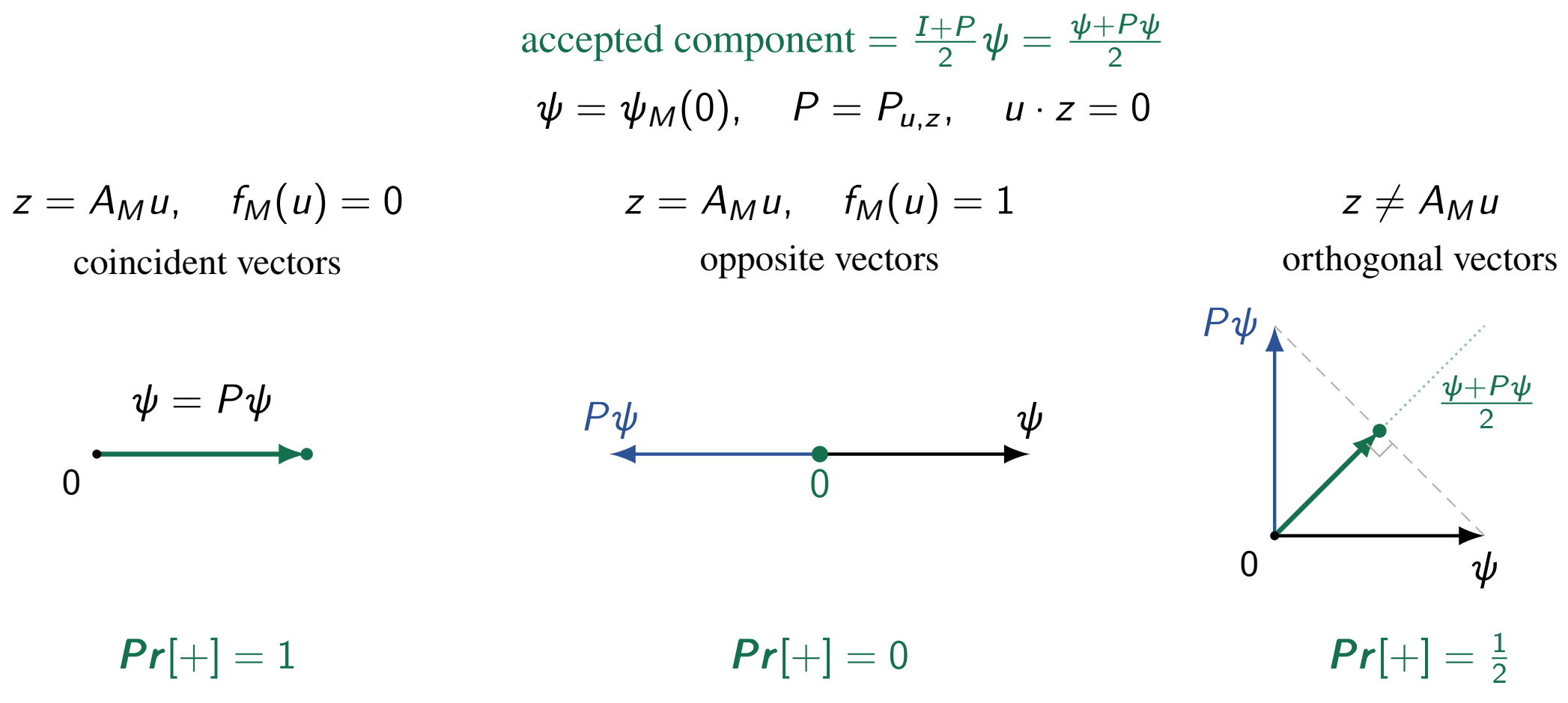

For $a\in\F^d$, define
\m{
\psi_M(a)\deq2^{-d/2}\sum_{x\in\F^d}(-1)^{f_M(x)+a\tm x}\bk{x}.
}
These vectors form an orthonormal basis.
Alice sends $\psi_M(0)$ to Bob, using $d$ qubits.
Bob rejects if $u\tm z=1$ (as $A_M$ is alternating).
Otherwise he measures using the projectors $(I\pm P_{u,z})/2$, where
\m{
P_{u,z}\bk{x}\deq(-1)^{z\tm x}\bk{x+u},
}
and accepts iff the outcome is $+1$.
For $u\tm z=0$, the operator $P_{u,z}$ is Hermitian, satisfies $P_{u,z}^2=I$ and maps each $\psi_M(a)$ to a basis vector or its negative:
\m{
f_M(x+u)+f_M(x)&=f_M(u)+(A_Mu)\tm x,\\
P_{u,z}\psi_M(a)
&=(-1)^{f_M(u)+a\tm u}\psi_M(a+A_Mu+z).
}
The accepting component is the midpoint of $\psi_M(0)$ and $P_{u,z}\psi_M(0)$, as shown in \figref{fig:oneway-geometry}.
Its squared norm is the acceptance probability:
\m[eq:acceptance]{
\PR{\txt{accept}}
&=\fr{1+\bk[\psi_M(0)]<P_{u,z}>{\psi_M(0)}}{2}\\
&=\fr{1+(-1)^{f_M(u)}\one[z=A_Mu]}{2}.
}

Repeating the test independently twice and accepting iff both tests pass gives one-sided error at most $1/4$, using $2d$ qubits.

\ssect{SMP versions}

Alice and Bob share $d$ EPR pairs.
Alice applies the phase $\bk{x}\mapsto(-1)^{f_M(x)}\bk{x}$ to her share and measures in the Hadamard basis.
She sends her $d$-bit outcome $a$ to the referee, leaving Bob with $\psi_M(a)$.
If $u\tm z=1$, Bob chooses $s\in\II$ uniformly at random; otherwise he measures using $(I\pm P_{u,z})/2$ and records the outcome $s$.
He sends $(u,s)$ to the referee.
The referee accepts iff $s=(-1)^{a\tm u}$.
For $u\tm z=0$,
\m{
\PR{\txt{accept}}[a]
&=\fr{1+(-1)^{a\tm u}\bk[\psi_M(a)]<P_{u,z}>{\psi_M(a)}}{2}\\
&=\fr{1+(-1)^{f_M(u)}\one[z=A_Mu]}{2},
}
which is the same probability as in~\eqref{eq:acceptance}.
For $u\tm z=1$, the acceptance probability is $1/2$.
Repeating twice independently, sending $u$ only once and accepting iff both runs accept gives one-sided error at most $1/4$, using $2d$ EPR pairs and $3d+2$ classical bits.

For quantum SMP, Bob sends his $2d$-bit classical input $(u,z)$ to the referee, who runs the one-way protocol on Alice's quantum message.

\ssect{Other bounds}

On $M=\diag{b}$, $z=0$, the function is the complement of inner product.
Cleve, van Dam, Nielsen and Tapp~\cite{CleveDNT} give an $\asOm{d}$ lower bound even for two-way quantum communication with arbitrary prior entanglement, matching the $\asO{d}$ protocols above.
Classical two-way communication has cost in $\asT{d}$: Bob sends $(u,z)$ and Alice returns the answer.
The classical one-way and SMP upper bounds follow by transmitting the inputs.

\sect[sec:classical]{The classical complexity}

We use the following estimate, proved in \sref{sec:moment}.
\nprp[prop:moment]{Uniform moments}
{
For any function $g:\F^d\times\F^d\to\II$, define
\m{
R_g(M)\deq\E[u\unin\F^d]{g\lrp{u,(M+M^{\mathsf T})u}(-1)^{u^{\mathsf T}Mu}}.
}
Then, for every even $p\ge2$,
\m{
\E[M\unin\F^{d\times d}]{\sz{R_g(M)}^p}\le2^{2p^2-\fr1{10}pd}.
}
}

Choose $M\unin\F^{d\times d}$, $u\unin\F^d$ and $z=A_Mu$.
Assume a fixed $c$-bit message, at a cost of at most one extra bit, and fix the randomness while retaining average success at least $2/3$.
After message $j\in[2^c]$, write Bob's answer on $(u,z)$ as $h_j(u,z)\in\II$, with $+1$ denoting acceptance.
The correct answer on this distribution is $(-1)^{f_M(u)}~=~(-1)^{u^{\mathsf T}Mu}$, so
\m{
\fr13
&~~\le~~\E[M]{\Max[j\in[2^c]]{\sz{\E[u]{h_j(u,A_Mu)\tm(-1)^{u^{\mathsf T}Mu}}}}}
~~=~~\E[M]{\Max[j\in[2^c]]{\sz{R_{h_j}(M)}}}\\
&~~\le~~\lrp{2^c \tm \E[j\unin[2^c],M]{\sz{R_{h_j}(M)}^p}}^{1/p}\\
&~~\le~~\lrp{2^c \tm 2^{-\fr1{10}pd+\asO{p^2}}}^{1/p}.
}
Taking logarithms gives $c\ge \fr1{10}pd-\asO{p^2}$; choosing an even $p\in\asT{d}$ with a sufficiently small proportionality constant yields $c\in\asOm{d^2}$.
Together with $\Rone\le\Rsmp$ and the bounds of \sref{sec:qua-com}, this proves \theoref{thm:main}.

\sect[sec:moment]{The moment estimate}

\prfstart[of \prpref{prop:moment}]
All averaged vector variables below are independent and uniform on $\F^d$.
For vector spaces, $\le$ denotes subspace inclusion.
The Fourier transform of $g$ in its second argument is
\m{
\hat g(u;s)\deq\E[z]{g(u,z)(-1)^{z\tm s}}.
}
Parseval gives
\m[eq:diagonal-norm]{
\sum_s\sz{\hat g(u;s)}^2=1.
}
In particular, $\Max[u,s]{\sz{\hat g(u;s)}}\le1$ and $\sum_{u,s}\sz{\hat g(u;s)}\le 2^{3d/2}$ by Cauchy--Schwarz.
Fourier inversion gives:
\m{
R_g(M)&=\E[u]{g(u,A_Mu)(-1)^{f_M(u)}}\\
&=2^d \tm \E[u,z]{g(u,z)(-1)^{f_M(u)}\E[s]{(-1)^{(A_Mu+z)\tm s}}}\\
&=2^d \tm \E[r,s]{\hat g(r+s;s)(-1)^{f_M(r)+f_M(s)}}.
}

Write $Y\deq(r_1\dc r_p,s_1\dc s_p)\in\F^{d\times2p}$.
In the expansion of $R_g(M)^p$, averaging over $M$ retains exactly the terms with $YY^{\mathsf T}=\sum_ir_ir_i^{\mathsf T}+\sum_is_is_i^{\mathsf T}=0$:
\m[eq:moment-expansion]{
\E[M]{\sz{R_g(M)}^p}
&=2^{-pd} \tm \sum_{YY^{\mathsf T}=0}\prod_i\hat g(r_i+s_i;s_i)\\
&\le2^{-pd} \tm \sum_{L\le\F^{2p}:\,L\sbseq L^\perp}Z_L,\\
Z_L&\deq\sum_{Y:\,\operatorname{rows}(Y)\sbseq L}\prod_i\sz{\hat g(r_i+s_i;s_i)}.
}
Taking absolute values and summing over all self-orthogonal subspaces gives the inequality, since $YY^{\mathsf T}=0$ makes the rowspace of $Y$ self-orthogonal.

Fix one such $L$, choose a generator matrix $G=(R\ S)$ whose rows form a basis of $L$ and put $k\deq\Dim{L}\le p$ and $m\deq\rank{R+S}$.
Greedily choose a maximal family of column pairs $(R_i,S_i)$ such that all selected columns are linearly independent and let $q$ be the number of selected pairs.
List the selected pairs first in each block and perform row operations; these preserve $Z_L$ and $m$.
Every matrix $Y$ with rows in $L$ then has the unique form
\m[C]{
Y=\underbrace{\lrp{X_1\ X_2\ X_3}}_{X\in\F^{d\times k}}
\underbrace{\lrp{\begin{array}{cc|cc}
I_q&*&0&*\\
0&*&I_q&*\\ \hline
0&R'&0&S'
\end{array}}}_{G=(R\ S)}
=\lrp{X_1\qd *\mid X_2\qd *}.
}
Here $X_1,X_2$ have $q$ columns each and $X_3$ has $k-2q$; the starred blocks are determined by $G$ and $X$.
We first note that
\m[eq:greedy-rank]{
2m\le p+3q.
}
Indeed, deleting the selected columns from the bottom $k-2q$ rows leaves a self-orthogonal matrix $(R'\ S')$.
By maximality, each pair $(R'_i,S'_i)$ is dependent, so
\m{
(R'+S')(R'+S')^{\mathsf T}=R'R'^{\mathsf T}+S'S'^{\mathsf T}=0.
}
Hence
\m{
m\le2q+\rank{R'+S'}\le2q+\fr{p-q}{2}.
}

There are two bounds for $Z_L$.
First, fixing $X_1,X_2$ leaves $2^{d(k-2q)}$ choices of $X_3$.
Discarding the other factors using $\sz{\hat g}\le1$ gives
\m{
Z_L\le 2^{d(k-2q)} \tm \lrp{\sum_{u,s}\sz{\hat g(u;s)}}^q
\le 2^{d(k-q/2)}.
}
Second, there are $2^{dk}$ coefficient matrices $X$ and $2^{dm}$ possible difference arrays $v\deq(r_i+s_i)_i=X(R+S)$.
Cauchy--Schwarz and~\eqref{eq:diagonal-norm} give
\m{
Z_L^2
&\le2^{dk} \tm \sum_{Y:\,\operatorname{rows}(Y)\sbseq L}\prod_i\sz{\hat g(r_i+s_i;s_i)}^2\\
&\le2^{dk} \tm \sum_v\prod_i\lrp{\sum_{s_i}\sz{\hat g(v_i;s_i)}^2}
=2^{d(k+m)}.
}
Cubing the first bound and multiplying by the second, using~\eqref{eq:greedy-rank} and $k\le p$, gives
\m{
Z_L^5&\le2^{d(4k+m-\fr32q)}\le2^{\fr92pd},\\
Z_L&\le2^{\fr9{10}pd}.
}

Every self-orthogonal $L$ has dimension at most $p$ and is the rowspace of some binary $p\times2p$ matrix.
There are $2^{2p^2}$ such matrices; hence \eqref{eq:moment-expansion} gives
\m{
\E[M]{\sz{R_g(M)}^p}\le2^{2p^2-\fr1{10}pd},
}
as required.
\prfend[\prpref{prop:moment}]

\bibstart
\small
\setlength{\itemsep}{0pt}

\bibitem{CleveDNT}
Richard Cleve, Wim van Dam, Michael Nielsen and Alain Tapp.
\newblock Quantum entanglement and the communication complexity of the inner product function.
\newblock \e{Theoretical Computer Science}, 486:11--19, 2013.

\bibend
\end{document}